\documentclass[a4paper,11pt,fleqn]{article}

\usepackage{amsmath,amssymb,amsthm,cite} 
\usepackage{rotating}

\newcommand{\todo}[1][\null]{\ensuremath{\clubsuit}}
\newcommand{\checked}[1][\null]{\ensuremath{\diamond}}
\newcommand{\noprint}[1]{}

\newcounter{tbn}

\newcounter{mcasenum}

\newtheorem{theorem}{Theorem}

\newtheorem{proposition}{Proposition}
\newtheorem*{proposition*}{Proposition}
{\theoremstyle{definition}

\newtheorem{example}{Example}

}

\begin{document}
\begin{center}
\par\noindent {\LARGE\bf
Enhanced group classification of Gardner equations\\ with time-dependent coefficients

\par}

{\vspace{5mm}\par\noindent\large Olena~Vaneeva$^{\dag 1}$, Oksana Kuriksha$^{\ddag 2}$ and Christodoulos Sophocleous$^{\S 3}$
\par\vspace{2mm}\par}
\end{center}

{\par\noindent\it\small
${}^\dag$\ Institute of Mathematics of NAS of Ukraine,
 3 Tereshchenkivska Str., 01601 Kyiv-4, Ukraine \\[1ex]
${}^\ddag$\  Petro Mohyla Black Sea State University, 10, 68 Desantnykiv Street, 54003 Mykolaiv, Ukraine\\[1ex]
${}^\S$\ Department of Mathematics and Statistics, University of Cyprus, Nicosia CY 1678, Cyprus
}

{\vspace{3mm}\par\noindent
$\phantom{{}^\dag{}\;}\ $E-mails: \it $^1$vaneeva@imath.kiev.ua, $^2$oksana.kuriksha@gmail.com, $^3$christod@ucy.ac.cy

\par}

{\vspace{5mm}\par\noindent\hspace*{5mm}\parbox{150mm}{\small
We classify the Lie symmetries of variable coefficient Gardner equations (called also the combined KdV-mKdV equations).
In contrast to the particular results  presented in [M.~Molati, M.P.~Ramollo, {\it Commun.\ Nonlinear Sci.\ Numer.\ Simulat.\ }{\bf 15} (2012), 1542--1548] we perform the exhaustive group classification. It is shown that the complete results can be achieved using either the gauging of arbitrary elements of the class  by the equivalence transformations or the method of mapping between classes.
As by-product of the second approach the complete group classification of a class of variable coefficient mKdV equations with forcing term is derived.
Advantages of the use of the generalized extended equivalence group in comparison with the usual one  are also discussed.

}\par\vspace{4mm}}

\section{Introduction}

Lie symmetry analysis proved itself as a powerful and algorithmic tool for studying differential
equations (DEs). In spite of its original
goal of finding exact solutions for DEs (especially for nonlinear ones) Lie symmetries  have been found useful in construction of conservation laws~\cite{Noether},
seeking fundamental solutions~\cite{CraddockPlaten2004}, solving initial and boundary value problems~\cite{Bluman&Anco2002}, construction of numerical solutions (see, e.g.,~\cite{VPCS2014}), study of complicated systems using invariant submodels~\cite{Ovsyannikov1994},
derivation of physically important models using the requirement of invariance under certain group of transformations (like, e.g., Galilei or Poincar\'e group)~\cite{FN}, etc.

One of the central problems of group analysis is the {\it group classification problem} that concerns not a single DE but a class of DEs (DE or a system of DEs that is parameterized by arbitrary elements being constants and/or functions). The solution of the problem
implies finding the Lie symmetry group admitted by any DE from a given class and deriving  all inequivalent values of arbitrary elements for which the corresponding DEs possess Lie symmetry extensions.

There is unceasing interest in solving group classification problems for various classes of DEs that are of current or potential interests for applications. Many such classes involve several arbitrary functions (variable coefficients), which often  makes their symmetry analysis difficult. To overcome these obstacles
a number of useful tools and notions were proposed.
These are, in particular, notions of generalized~\cite{Meleshko1994} and extended~\cite{mogran} equivalence groups, admissible \cite{popo2010a} (form-preserving~\cite{Kingston&Sophocleous1998}, allowed~\cite{Winternitz92}) transformations, equivalence groupoid~\cite{Popovych&Bihlo2012}, normalized class of DEs~\cite{popo2010a}, contractions of equations and corresponding symmetries~\cite{monstr2,VPS2012}; the method of furcate split~\cite{Popovych&Ivanova2004NVCDCEs}, the partition of a non-normalized class into normalized subclasses~\cite{BihloPopovychJMP2012,popo2010a}, the method of mapping between classes~\cite{VPS2009}.

Nevertheless, there is still a number of works where such tools are neglected and only particular results are derived instead of complete classifications.
One of such works is the recent classification
of the variable coefficient Gardner equations
\begin{gather}\label{eq_gGardner}
u_t+k(t)uu_x+f(t)u^2u_x+g(t)u_{xxx}=0, \qquad f g\neq0,
\end{gather}
presented in~\cite{Molati&Ramollo2012}. Here
$k$, $f$, and $g$ are smooth functions of the variable $t.$

In the present paper we achieve the exhaustive classification using the groups of equivalence transformations of class~\eqref{eq_gGardner} that are found in Section~2.
We show that even the use of usual equivalence group allows one to get the complete result. At the same time utilizing wider generalized extended equivalence group
provides more simplification and therefore is preferable. This is illustrated in the process of finding Lie symmetries
of equations~\eqref{eq_gGardner} in Section~3.1. We check the obtained results using the alternative method of mapping between classes in Section~3.2. As by-product of the latter approach the exhaustive Lie symmetry classification of the related class of variable coefficient mKdV equations with forcing term is derived. A discussion on optimal choice of the method and a brief comparison of the obtained results with those presented in~\cite{Molati&Ramollo2012} are given in the conclusion.

\section{Equivalence transformations}
Firstly, we search for nondegenerate point transformations, that preserve the differential structure of the  class~\eqref{eq_gGardner} and change only its arbitrary elements. They are called {\it equivalence transformations} and form a group. There are several kinds of equivalence groups. The {\it usual equivalence group}, used by Ovsiannikov  for solving group classification problems since late 50's,  consists of the nondegenerate point transformations
of the independent and dependent variables and of the arbitrary elements of the class,
where transformations for independent and dependent variables do not involve arbitrary elements of class~\cite{Ovsiannikov1982}.
In 1994 Meleshko suggested to consider the {\it generalized equivalence group}, where transformations of variables of given DEs explicitly depend on arbitrary elements~\cite{Meleshko1994,Meleshko1996}.
The attribute {\it extended} for equivalence groups was proposed to distinguish those equivalence groups  whose transformations
 include nonlocalities with respect to arbitrary elements
 (e.g., if new arbitrary elements  are expressed via integrals of old ones)~\cite{mogran}.

 Given a class of DEs, if we consider the set of triples each of which consists of two fixed equations from class and a point transformation linking them (such triples are called {\it admissible transformations} and the  entire set of them is called {\it equivalence groupoid}~\cite{Popovych&Bihlo2012}), then equivalence transformations generate a subset in this set. If the  set of  admissible transformations is generated by the equivalence group of a class,
then this class is  called {\it normalized}~\cite{popo2010a}. The normalization property has appeared to be rather important in group analysis. Thus,  algebraic method of group classification guarantees the complete
 result for normalized classes only~\cite{BihloPopovychJMP2012,popo2010a}. It was shown also that a reasonable way of solving group classification problems in classes that are not normalized is the partition of such classes into normalized subclasses~\cite{popo2010a,VPS2012}.

 Generators of one-parameter subgroups of the equivalence group can be found by the Lie infinitesimal method, whereas the direct method~\cite{king1991c,Kingston&Sophocleous1998} allows one to find the entire equivalence group including even discrete equivalence transformations and therefore this technique is preferable.
A very useful feature of normalized classes is that the equivalence groups for their subclasses, singled out by setting certain restrictions on arbitrary elements, are subgroups of the equivalence group of the entire class.
We will use this property to derive the equivalence group of the class~\eqref{eq_gGardner}.

It was proven in~\cite{Popovych&Vaneeva2010} that the more general class of mKdV-like equations
\begin{equation}\label{EqvcmKdV}
u_t+f(t)u^2u_x+g(t)u_{xxx}+h(t)u+(p(t)+q(t)x)u_x+k(t)uu_x+l(t)=0,
\end{equation}
where all the parameters are arbitrary smooth functions of~$t$, $fg\ne0$,
is  normalized in the usual sense. In other words, all point transformations that connect equations from this class are induced by transformations from its usual equivalence group.
This group consists of the transformations
\[\tilde t=\alpha(t),\quad
\tilde x=\beta(t)x+\gamma(t),\quad
\tilde u=\theta(t)u+\psi(t),\]
where $\alpha$, $\beta$, $\gamma$, $\theta$ and $\psi$ run through the set of smooth functions of~$t$ and $\alpha_t\beta\theta\ne0$.
The arbitrary elements of~\eqref{EqvcmKdV} are transformed by the formulae~\cite{Popovych&Vaneeva2010}:
\begin{gather*}
\tilde f=\frac{\beta}{\alpha_t\theta^2}f, \quad
\tilde g=\frac{\beta^3}{\alpha_t}g, \quad
\tilde k=\frac\beta{\alpha_t\theta}\left(k-2\frac\psi\theta f\right), \quad
\tilde l=\frac1{\alpha_t}\left(\theta l-\psi h-\psi_t+\psi\frac{\theta_t}\theta\right),\\
\tilde h=\frac1{\alpha_t}\left(h-\frac{\theta_t}\theta\right), \quad
\tilde p=\frac1{\alpha_t}\left(\beta p-\gamma q+\beta\frac{\psi^2}{\theta^2} f-\beta\frac\psi\theta k+\gamma_t-\gamma\frac{\beta_t}\beta\right), \quad
\tilde q=\frac1{\alpha_t}\left(q+\frac{\beta_t}\beta\right)\!.
\end{gather*}
As class~\eqref{EqvcmKdV} is normalized we are able to derive all admissible transformations of class~\eqref{eq_gGardner} simply by setting  $\tilde l=l=\tilde h=h=\tilde p=p=
\tilde q=q=0$ in the latter formulas. Note that for classes that are not normalized this may lead to incomplete results. As a result we obtain the equations $\beta_t=\theta_t=\psi_t=0$ and $\beta\psi(\psi f-\theta k)+\gamma_t\theta^2=0.$ Their solution is
$\beta=\delta_1$, $\theta=\delta_2$, $\psi=\delta_3,$ and $\gamma=\delta_1\delta_3\delta_2^{\,-2}\int(\delta_2 k -\delta_3 f) {\rm d}t+\delta_4$, where $\delta_i$, $i=1,\dots,4$, are arbitrary constants with $\delta_1\delta_2\neq0.$
There is no additional constraint for the function $\alpha$, therefore, it is an arbitrary smooth function with $\alpha_t\neq0.$ The following two assertions are true.

\begin{theorem} The generalized extended equivalence group~$\hat G^{\sim}$ of class~\eqref{eq_gGardner} is formed by the transformations
\begin{gather*}
\tilde t=\alpha(t),\quad \tilde x=\delta_1 x+\frac{\delta_1\delta_3}{\delta_2^{\,2} }\int(\delta_2 k(t)-\delta_3 f(t)) {\rm d}t+\delta_4,\quad
\tilde u=\delta_2 u+\delta_3, \\[1ex]
\tilde k(\tilde t)=\frac{\delta_1}{\delta_2 \alpha_t}\left(k(t)-2\frac{\delta_3}{\delta_2}f(t)\right),\quad\tilde f(\tilde t)=\frac{\delta_1}{\delta_2^{\,2} \alpha_t}f(t),\quad\tilde g(\tilde t)=\dfrac{\delta_1^{\,3}}{\alpha_t} g(t),
\end{gather*}
where  $\delta_i,$ $i=1,\dots,4$, are arbitrary constants with
$\delta_1\delta_2\not=0$, $\alpha$ is an arbitrary smooth function with $\alpha_t\neq0.$

The usual equivalence group~$G^{\sim}$ of class~\eqref{eq_gGardner} consists of the above transformations with $\delta_3=0.$
\end{theorem}
\begin{proposition}
The entire set of admissible transformations (equivalence groupoid) of class~\eqref{eq_gGardner}   is generated by the transformations from the group~$\hat G^{\sim}$. Class~\eqref{eq_gGardner} is normalized in the generalized extended sense.
\end{proposition}

Thus, there are no other point transformations between equations from class~\eqref{eq_gGardner} than those transformations from the group~$\hat G^{\sim}$.
To deduce which variable coefficient equations of the form~\eqref{eq_gGardner} is reducible to their constant coefficient counterparts we assume $\tilde k$ and $\tilde f$ are constant in the transformation  components for arbitrary elements in~$\hat G^{\sim}$, this results in the following statement.

\begin{proposition}
A variable coefficient equation from class~\eqref{eq_gGardner} is reducible to constant coefficient equation from the same class if and only if
the coefficients $f,$ $g$ and $k$ satisfy the conditions
\begin{equation*}\label{criterion1}
(f/k)_t=(g/k)_t=0.
\end{equation*}
\end{proposition}

As there is one arbitrary function, $\alpha(t)$, in the transformations from the group~$\hat G^{\sim}$, we can set one of the arbitrary elements  of class~\eqref{eq_gGardner} to a nonzero constant value. We choose
the gauging $g=1$ and perform it using the transformation
\begin{equation}\label{tr}\tilde t=\int\! g(t){\rm d}t, \quad \tilde x=x,\quad\tilde u=u.\end{equation} Then, any equation from the class~\eqref{eq_gGardner} is mapped to one from its subclass singled out by the condition  $g=1$. Old forms of the arbitrary elements are connected with  new ones  via the formulae $\tilde k=k/g$ and $\tilde f=f/g$.

Note that the most general form of transformation that maps an equation from class~\eqref{eq_gGardner} to another equation  from the same class with $g=1$ is
\begin{equation}\label{gen_tr}
\tilde t=\delta_1^{\,3}\int\! g(t){\rm d}t+\delta_0,\quad \tilde x=\delta_1 x+\frac{\delta_1\delta_3}{\delta_2^{\,2} }\int(\delta_2 k(t)-\delta_3 f(t)) {\rm d}t+\delta_4,\quad
\tilde u=\delta_2 u+\delta_3,
\end{equation}
where $\delta_i$, $i=0,\dots,4,$ are constants with $\delta_1\delta_2\neq0$.

Without loss of generality, we can restrict ourselves to the study of the class
\begin{gather}\label{eq_Gardner}
u_t+k(t)uu_x+f(t)u^2u_x+u_{xxx}=0,
\end{gather}
since all results on symmetries, conservation laws, classical solutions and other related objects can be found for equations~\eqref{eq_gGardner}
using the similar results derived for equations~\eqref{eq_Gardner}.

As class~\eqref{eq_gGardner} is normalized in the generalized extended sense, in order to derive the equivalence group for its subclass with $g=1$ it is enough to set $\tilde g=g=1$ in the transformations  from the group~$\hat G^{\sim}$ presented in Theorem~1. This leads to the equation for $\alpha$: $\alpha_t=\delta_1^{\,3},$  resulting in $\alpha=\delta_1^{\,3}t+\delta_0,$ where $\delta_0$ is an arbitrary constant.
The following statement is true.

\begin{theorem}The generalized extended equivalence group~$\hat G^{\sim}_1$ of class~\eqref{eq_Gardner} comprises  the transformations
\begin{gather*}\label{tr_equiv_cor2}
\tilde t=\delta_1^{\,3}t+\delta_0,\quad \tilde x=\delta_1 x+\frac{\delta_1\delta_3}{\delta_2^{\,2}}\int(\delta_2 k(t)-\delta_3 f(t)) {\rm d}t+\delta_4,\quad
\tilde u=\delta_2 u+\delta_3, \\[1ex]
\tilde k(\tilde t)=\frac{\delta_2 k(t)-2\delta_3 f(t)}{\delta_1^{\,2}\delta_2^{\,2}},\quad\tilde f(\tilde t)=\frac{f(t)}{\delta_1^{\,2}\delta_2^{\,2}},
\end{gather*}
where  $\delta_i$, $i=0,\dots,4,$ are arbitrary constants with
$\delta_1\delta_2\not=0$.

The usual equivalence group~$G^{\sim}_1$ of class~\eqref{eq_Gardner} consists of the above transformations with $\delta_3=0.$
\end{theorem}
We note that class~\eqref{eq_Gardner} is normalized in the generalized extended sense. From Proposition~2 we deduce that there are no variable coefficient equations~\eqref{eq_Gardner} that are reducible to constant coefficient equations from the same class by point transformations.

In the next section we demonstrate usage of the found equivalence transformations  in the process of group classification. Simplifications by usual and generalized equivalence groups will be compared.

\section{Classification of Lie symmetries}
There are two main approaches in modern group analysis for solving group classification problems: the algebraic method based on
the subgroup analysis of the corresponding equivalence group~\cite{Basarab-Horwath&Lahno&Zhdanov2001,BihloPopovychJMP2012} and
the ``direct'' approach based on integration of determining equations (an overdetermining  system of linear PDEs) implied by the infinitesimal invariance criterion~\cite{Ovsiannikov1982}. If a class of DEs is parameterized by several arbitrary elements, then group classification problem may appear to be too complicated to be solved completely. To solve group classification problems for such classes new techniques based on usage of point transformations were developed recently. These are, in particular, the gauging of arbitrary elements by equivalence transformations (i.e., reducing of a class to its subclass with fewer number of arbitrary elements) and the method of mapping between classes~\cite{VPS2009}. We will perform the group classification of class~\eqref{eq_gGardner} using the gauging of arbitrary elements by equivalence transformations (Section~3.1) and will verify the results utilizing the method of mapping between classes (Section~3.2).

\subsection{Group classification using equivalence transformations}
In Section~2 we have determined the widest equivalence group of the class~\eqref{eq_gGardner} that appeared to be generalized extended one.
The gauging $g=1$ was performed using transformation~\eqref{tr}. In such a way the group classification problem for the the class~\eqref{eq_gGardner} was reduced to
the group classification problem for its subclass~\eqref{eq_Gardner}. More precisely,
the group classification of class~\eqref{eq_Gardner} up to $\hat G^{\sim}_1$-equivalence coincides with
the group classification of class~\eqref{eq_gGardner} up to $\hat G^{\sim}$-equivalence.
We carry out the group classification of the class~\eqref{eq_Gardner}
using the  classical algorithm~\cite{Olver1986,Ovsiannikov1982}.
Namely, we
search for symmetry operators of the form \[Q=\tau(t,x,u)\partial_t+\xi(t,x,u)\partial_x+\eta(t,x,u)\partial_u\]  generating one-parameter Lie groups
of transformations that leave equations~\eqref{eq_Gardner} invariant.
We require that the action of the third prolongation $Q^{(3)}$ of the operator~$Q$ on left-hand side of~\eqref{eq_Gardner} vanishes identically modulo equation~\eqref{eq_Gardner},
\begin{equation}\label{c2}
Q^{(3)}\{u_t+k(t)uu_x+f(t)u^2u_x+u_{xxx}\}|_{u_t=-k(t)uu_x-f(t)u^2u_x-u_{xxx}}=0.
\end{equation}
Here
 $Q^{(3)}=Q+\eta^t\partial_{u_t}+\eta^x\partial_{u_x}+\eta^{xxx}\partial_{u_{xxx}}$, where
\begin{eqnarray*}
&\eta^t=D_t(\eta)-u_tD_t(\tau)-u_xD_t(\xi),&
\eta^x=D_x(\eta)-u_tD_x(\tau)-u_xD_x(\xi),\\
&\eta^{xx}=D_x(\eta^x)-u_{tx}D_x(\tau)-u_{xx}D_x(\xi),&
\eta^{xxx}=D_x(\eta^{xx})-u_{txx}D_x(\tau)-u_{xxx}D_x(\xi),
\end{eqnarray*}
$D_t=\partial_t+u_t\partial_{u}+u_{tt}\partial_{u_t}+u_{tx}\partial_{u_x}+\dots{}$ and
$D_x=\partial_x+u_x\partial_{u}+u_{tx}\partial_{u_t}+u_{xx}\partial_{u_x}+\dots{}$
are the total derivatives with respect to~$t$ and~$x$, respectively.

The infinitesimal invariance criterion~\eqref{c2} implies the determining equations, simplest of which result in
\[
\tau=\tau(t),\quad
\xi=\xi(t,x), \quad
\eta=\eta^1(t,x)u+\eta^0(t,x),
\]
where $\tau$, $\xi$, $\eta^1$ and $\eta^0$ are arbitrary smooth functions of their variables. This was verified using the  MAPLE-based  GeM software package~\cite{Cheviakov}.
Then the rest of the determining equations are
\begin{gather*}
\eta^1_x=\xi_{xx},\quad
\tau _t =3\xi_x,\\
\eta^1_xfu^3+(\eta^1_xk+\eta^0_xf)u^2+(\eta^1_t+\eta^1_{xxx}+\eta^0_xk)u+\eta^0_t+\eta^0_{xxx}=0,\\
(\tau f_t+(\tau_t-\xi_x+2\eta^1)f)u^2+(\tau k_t+2\eta^0f+(\tau_t-\xi_x+\eta^1)k)u+\eta^0k+3\eta^1_{xx}-\xi_{xxx}-\xi_t=0.
\end{gather*}
As the functions $\tau$, $\xi$, $\eta^1$ and $\eta^0$ do not depend on $u$, we can split the third and the fourth determining equations with respect to this variable. As a result we get the system
\begin{gather}\label{deteq1}
\eta^1_x=\eta^0_x=\eta^1_t=\eta^0_t=\xi_{xx}=0,\quad
\tau _t =3\xi_x,\quad \eta^0k=\xi_t,\\\label{deteq2}
\tau f_t+(\tau_t-\xi_x+2\eta^1)f=0,\quad \tau k_t+(\tau_t-\xi_x+\eta^1)k+2\eta^0f=0.
\end{gather}
The integration of~\eqref{deteq1} leads to  $\tau= 3c_1 t+c_0$, $\xi = c_1x+c_2+c_4\int k(t){\rm d}t$, $\eta^1=c_3$ and $\eta^0=c_4$,
where $c_i,$ $i=0,\dots,4$, are arbitrary constants.
Therefore, the general form of the infinitesimal generator admitted by equations~\eqref{eq_gGardner} is given by
\begin{equation}\label{op_Q}\textstyle
Q=(3c_1 t+c_0)\partial_t+( c_1x+c_2+c_4\int\! k(t){\rm d}t)\partial_x+(c_3u+c_4)\partial_u.
\end{equation}
Using equations \eqref{deteq2}, the classifying equations involving the arbitrary  functions $f$ and $k$ as well as the residuary uncertainties in the coefficients of the operator $Q$, given by~\eqref{op_Q}, are
\begin{equation}\label{clas_eq}
(3c_1 t+c_0) f_t=-2(c_1+c_3)f,\quad (3c_1 t+c_0) k_t=-(2c_1+c_3)k-2c_4f.
\end{equation}
In order to find the Lie invariance algebra admitted by any equation from class~\eqref{eq_Gardner} (so-called kernel algebra),
 we split~\eqref{clas_eq} with respect to $f$, $k$ and their derivatives.
This results in $c_0=c_1=c_3=c_4=0$ and therefore  $Q=c_2\partial_x$. Thus, the kernel algebra~$A^{\rm ker}$ of maximal Lie invariance algebras~$A^{\rm max}$ of equations from class~\eqref{eq_Gardner} is the one-dimensional algebra $\langle\partial_x\rangle$.
To get possible extensions of $A^{\rm ker}$ we consider~\eqref{clas_eq} not as two identities but as a system of first-order ODEs  on $f$ and $k$, that is of the form
\begin{gather*}
(at+b)f_t=cf, \quad (at+b)k_t=\left(\frac{c}{2}-\frac{a}{3}\right)k+df,
\end{gather*}
where $a,$ $b$, $c$ and $d$ are arbitrary constants with $a^2+b^2\neq0$.
The system  should be integrated up to the chosen equivalence.

The equivalence transformations from the groups $G^{\sim}_{1}$  and $\hat G^{\sim}_{1}$ combined with the multiplication on the nonzero arbitrary constant $\nu$ act on the coefficients $a,$ $b$, $c$ and $d$ of the above system in the following way
\begin{gather}\nonumber
G^{\sim}_{1}\colon\quad\tilde a=\nu a, \quad \tilde b=\nu\left(\delta_1^3b-\delta_0 a\right), \quad \tilde c=\nu c, \quad \tilde d=\nu \delta_2 d,\\\label{abcd_tr}
\hat G^{\sim}_{1}\colon\quad\tilde a=\nu a, \quad \tilde b=\nu\left(\delta_1^3b-\delta_0a\right), \quad \tilde c=\nu c, \quad \tilde d=\nu \left(\delta_2 d-\delta_3 c-\tfrac{2}{3}\delta_3 a\right).
\end{gather}

The transformations from the groups $G^{\sim}_{1}$  and $\hat G^{\sim}_{1}$ change the coefficients $a$, $b$ and~$c$ equally. In both cases
there are three cases of inequivalent  triples $(a,b,c)$ to be considered: I. $(1,0,\rho)$, II. $(0,1,1)$, and III. $(0,1,0)$, where $\rho$ is an arbitrary constant. Therefore, the three inequivalent systems of ODEs should be solved
\begin{eqnarray*}\label{I}
\phantom{II}{\rm I}.\quad& tf_t=\rho f,\quad& tk_t=\tfrac{3\rho-2}6k+\tilde df,\\\label{II}
\phantom{I}{\rm II}.\quad& f_t=f,\quad& \phantom{t}k_t=\tfrac12k+\tilde df,\\\label{III}
{\rm III.}\quad& f_t=0,\quad&\phantom{t}k_t=\tilde df,
\end{eqnarray*}
where $\tilde d$ is an arbitrary constant. The general solutions of~these systems are the following:
\begin{eqnarray*}
{\rm I.1}. &  f=\lambda_1t^\rho,\quad&  k=\lambda_2 t^{\frac{3\rho-2}{6}}+ \tfrac{6\lambda_1\tilde d}{3\rho+2} t^\rho,\\
{\rm I.2}. &  f=\lambda_1t^{-\frac23},\quad&  k=\lambda_2 t^{-\frac23}+\lambda_1\tilde d \ln|t|\,t^{-\frac23},\\
{\rm II}.&  f=\lambda_1e^t,\quad&  k=\lambda_2 e^{\frac12t}+2\lambda_1\tilde d e^t,\\\label{solIII}
{\rm III.}&  f=\lambda_1,\quad& k=\lambda_2+\lambda_1\tilde d\, t,
\end{eqnarray*}
where $\lambda_1$ and $\lambda_2$ are arbitrary constants with $\lambda_1\neq0$. In Case ${\rm I.1}$ $\rho\neq-2/3$, if $\tilde d\neq0$.

Now we will demonstrate how the choice of the equivalence group implies the result of group classification. It is easy to see from~\eqref{abcd_tr} that, if $c\neq-2/3\,a$, the constant $\tilde d$ can be set equal to zero by the transformations from the equivalence group~$\hat G^{\sim}_{1}$ with $\delta_3\neq0.$ The usual equivalence group~$G^{\sim}_{1}$ does not provide such simplification.
If we restrict ourselves by usage of the usual equivalence group~$G^{\sim}_{1}$, then  the following gauging of constants involved in the solutions for $f$ and $k$ can be made.
If $\tilde d\neq0$, we can scale it in such a way that the coefficients involving $\tilde d$ in the expressions for $k$ will be equal to unity.
That is why, this coefficient is denoted as $\delta$ in Table~1, where $\delta\in\{0,1\}$. Using scaling of $t$ we can also set $\lambda_2=0$ in Case ${\rm I.2}$. Case ${\rm III}$ is split into two subcases: if $\tilde d\neq0$, then $k$ can be reduced to $t$ by equivalence transformations
and, if $\tilde d=0$, then $k\in\{0,1\}$ (we denote the latter subcase as IV).

Now we substitute all inequivalent values of $f$ and $k$ into~\eqref{clas_eq} and find the values of $c_i,$ $i=0,\dots,4,$ and therefore the corresponding
forms of operator~\eqref{op_Q}. The results are presented in Table~1. As operator $Q$ involves the integral of $k$ in Case ${\rm I.1}$ the forms of $Q$ for $\rho=-1$ and $\rho=-4/3$ differ from ones for other values of $\rho$, that is why these cases are presented separately in Cases ${\rm I.3}$ and ${\rm I.4}$, respectively. In all the cases except IV ($k$ is a constant) the maximal Lie invariance algebras are two-dimensional, whereas in Case IV $A^{\rm max}$ is three-dimensional.

\begin{table}[t!]\small \renewcommand{\arraystretch}{1.8}
\centering
\setcounter{tbn}{-1}
\refstepcounter{table}\label{TableLieSym2}
\textbf{Table~\thetable.}
The group classification of class~\eqref{eq_gGardner} up to $G^\sim$-equivalence.
\\[2ex]
\begin{tabular}{|c|c|c|l|}
\hline
no.&$f(t)$&$k(t)$&\hfil Basis of $A^{\max}$ \\
\hline
0&$\forall$&$\forall$&$\partial_x$\\
\hline
I.1&
$\lambda_1 t^\rho$&$\lambda_2 t^{\frac{3\rho-2}{6}}+\delta  t^\rho$&$\partial_x,\,3t\partial_t+\left(x-\varkappa\frac{3\rho+2}{2}\left(\frac{6\lambda_2}{3\rho+4} t^{\frac{3\rho+4}{6}}+
\frac{\delta }{\rho+1} t^{\rho+1}\right)\right)\partial_x-\frac{3\rho+2}{2} \left(u+\varkappa\right)\partial_u$\\
\hline
I.2&
$\lambda_1 t^{-\frac23}$&$t^{-\frac23}\ln|t|$&$\partial_x,\,2\lambda_1t\partial_t+\left(\frac23\lambda_1x-3\left(\ln|t|-3 \right)t^{\frac{1}{3}}\right)\partial_x-\partial_u$\\
\hline
I.3&
$\lambda_1 t^{-1}$&$\lambda_2 t^{-\frac56}+\delta  t^{-1}$&$\partial_x,\,3t\partial_t+\left(x+\varkappa\left(3\lambda_2 t^{\frac{1}{6}}+
\frac12\delta  \ln{|t|}\right)\right)\partial_x+ \frac1{2}\left({u}+\varkappa\right)\partial_u$\\
\hline
I.4&
$\lambda_1 t^{-\frac43}$&$\lambda_2 t^{-1}+\delta  t^{-\frac43}$&$\partial_x,\,3t\partial_t+\left(x+\varkappa\left(\lambda_2 \ln{|t|}-3\delta  t^{-\frac{1}{3}}\right)\right)\partial_x+\left(u+\varkappa\right)\partial_u$\\
\hline
II&$\lambda_1 e^{t}$&$\lambda_2 e^{\frac t2}+\delta  e^t$&
$\partial_x,\,2\partial_t-\varkappa\left(2\lambda_2 e^{\frac t2}+\delta  e^t\right)\partial_x-\left(u+\varkappa\right)\partial_u$\\
\hline
III &$\lambda_1 $&$t$&
$\partial_x,\,2\lambda_1\partial_t-\frac12 t^2\partial_x-\partial_u$\\
\hline
IV&$\varepsilon$&$\delta $&
$\partial_x,\,\partial_t,\, 3t\partial_t+\left(x-\varkappa\,\delta t\right)\!\partial_x- \left(u+\varkappa\right)\partial_u$\\
\hline
\end{tabular}
\\[2ex]
\parbox{150mm}{Here $g=1\bmod G^\sim$; $\lambda_i,$ $i=1,2$, and $\rho$ are arbitrary constants with $\lambda_1\neq0$, $\rho\neq-\frac43,-1$; $\delta\in\{0,1\}\bmod G^\sim,$ $\varepsilon=\pm1\bmod G^\sim$, and $\varkappa=\frac12{\delta }/{\lambda_1}.$ In Case I.1 $(\rho,\delta)\neq(0,0)$.}
\end{table}
\begin{table}[t!]\small \renewcommand{\arraystretch}{1.8}
\centering
\setcounter{tbn}{-1}
\refstepcounter{table}\label{TableLieSym1}
\textbf{Table~\thetable.}
The group classification of class~\eqref{eq_gGardner} up to $\hat G^\sim$-equivalence.
\\[2ex]
\begin{tabular}{|c|c|c|l|}
\hline
no.&$f(t)$&$k(t)$&\hfil Basis of $A^{\max}$ \\
\hline
0&$\forall$&$\forall$&$\partial_x$\\
\hline
I.1&
$\lambda_1 t^\rho$&$\delta t^{\frac{3\rho-2}{6}}$&$\partial_x,\,3t\partial_t+x\partial_x-\frac{3\rho+2}{2} u\partial_u$\\
\hline
I.2&
$\lambda_1 t^{-\frac23}$&$t^{-\frac23}\ln|t|$&$\partial_x,\,2\lambda_1t\partial_t+\left(\frac23\lambda_1x-3\left(\ln|t|-3 \right)t^{\frac{1}{3}}\right)\partial_x-\partial_u$\\
\hline
II&$\lambda_1 e^{t}$&$\delta e^{\frac t2}$&
$\partial_x,\,2\partial_t-u\partial_u$\\
\hline
III &$\lambda_1 $&$t$&
$\partial_x,\,2\lambda_1\partial_t-\frac12 t^2\partial_x-\partial_u$\\
\hline
IV&$\varepsilon$&$0$&
$\partial_x,\,\partial_t,\, 3t\partial_t+x\partial_x- u\partial_u$\\
\hline

\end{tabular}
\\[2ex]
\parbox{150mm}{Here $g=1\bmod \hat G^\sim$;  $\lambda_1$ and $\rho$ are arbitrary constants with $\lambda_1\neq0$, $\delta\in\{0,1\}\bmod \hat G^\sim$, and $\varepsilon=\pm1\bmod G^\sim$. In Case~I.1 $(\rho,\delta)\neq(0,0)$.}
\end{table}

Now we consider the simplification by the transformations from the generalized extended equivalence group~$\hat G^\sim_1.$
Then in Cases ${\rm I.1}$  and ${\rm II}$ the constant $\tilde d$ can be scaled equal to zero, therefore, the function $k$ in these cases will take the form
$\lambda_2 t^{\frac{3\rho-2}{6}}$ and $\lambda_2 e^{\frac12t}$, respectively. Moreover, $\lambda_2$ can be set to $\delta\in\{0,1\}$ by scaling of $u.$ In Case IV any constant value of $k$ can be set equal to zero.
The constants in other cases are scaled in the same way as using the usual equivalence group. The results are presented in Table~2.

\looseness=-1
It is easy to see that
both forms of arbitrary elements and the corresponding symmetry generators are simpler in the case of usage of the group~$\hat G^\sim_1.$ The additional advantage is that  there is no partition of Case~${\rm I.1}$ for $\rho=-1$ and $\rho=-4/3$. Obviously the use of the widest (in our case generalized extended) equivalence group is preferable for solving  group classification problems.

We have proved the following statement.

\begin{sidewaystable} \small\renewcommand{\arraystretch}{2.3}
\begin{center}
\setcounter{tbn}{-1}
\refstepcounter{table}\label{TableLieSym3}
\textbf{Table~\thetable.}
The complete list of Lie symmetry extensions for the class~\eqref{eq_gGardner}.
\\[2ex]
\begin{tabular}{|c|c|c|l|}
\hline
no.&$f(t)$&$k(t)$&\hfil Basis of $A^{\max}$ \\
\hline
0&$\forall$&$\forall$&$\partial_x$\\
\hline
I.1&
$\lambda_1g (\alpha T+\beta)^\rho$&$\lambda_2g (\alpha T+\beta)^{\frac{3\rho-2}{6}}+\lambda_3g (\alpha T+\beta)^\rho$&$\partial_x,\,\,3 \dfrac {\alpha T+\beta}{g}\partial_t+\left(\alpha x-\varkappa\frac{3\rho+2}{2}\left(\frac{6\lambda_2}{3\rho+4} (\alpha T+\beta)^{\frac{3\rho+4}{6}}+
\frac{\lambda_3}{\rho+1} (\alpha T+\beta)^{\rho+1}\right)\right)\partial_x-\frac{3\rho+2}{2}\alpha  \left(u+\varkappa\right)\partial_u$\\
\hline
I.2&
$\lambda_1g (\alpha T+\beta)^{-\frac23}$&$(\lambda_2+\lambda_3 \ln|\alpha T+\beta|)g(\alpha T+\beta)^{-\frac23}$&$\partial_x,\,\dfrac{\alpha T+\beta}{g}\partial_t+\left(\frac13\alpha x-3\varkappa\left(\lambda_3 \ln{|\alpha T+\beta|}+\lambda_2-3\lambda_3 \right)(\alpha T+\beta)^{\frac{1}{3}}\right)\partial_x-\varkappa\alpha \partial_u$\\
\hline
I.3&
$\lambda_1 g (\alpha T+\beta)^{-1}$&$\lambda_2 g (\alpha T+\beta)^{-\frac56}+\lambda_3 g (\alpha T+\beta)^{-1}$&$\partial_x,\,\,3 \dfrac {\alpha T+\beta}{ g}\partial_t+\left(\alpha x+\varkappa\left(3\lambda_2 (\alpha T+\beta)^{\frac{1}{6}}+
\frac12\lambda_3 \ln{|\alpha T+\beta|}\right)\right)\partial_x+ \frac1{2}\alpha \left({u}+\varkappa\right)\partial_u$\\
\hline
I.4&
$\lambda_1 g (\alpha T+\beta)^{-\frac43}$&$\lambda_2 g (\alpha T+\beta)^{-1}+\lambda_3 g (\alpha T+\beta)^{-\frac43}$&$\partial_x,\,\,3 \dfrac {\alpha T+\beta}{g}\partial_t+\left(\alpha x+\varkappa\left(\lambda_2 \ln{|\alpha T+\beta|}-3\lambda_3 (\alpha T+\beta)^{-\frac{1}{3}}\right)\right)\partial_x+\alpha \left(u+\varkappa\right)\partial_u$\\
\hline
II&$\lambda_1 g e^{\alpha T}$&$\lambda_2 g e^{\frac12{\alpha T}}+\lambda_3 g e^{\alpha T}$&
$\partial_x,\,\,\dfrac 2g\partial_t-\varkappa\left(2\lambda_2 e^{\frac12{\alpha T}}+\lambda_3 e^{\alpha T}\right)\partial_x-\alpha\left(u+\varkappa\right)\partial_u$\\
\hline
III &$\lambda_1g $&$\lambda_2 g+\lambda_3gT$&
$\partial_x,\,\,\dfrac 1g\partial_t-\varkappa\, T\left(\frac12{\lambda_3}T+\lambda_2\right) \partial_x- \varkappa\,\partial_u $\\
\hline
IV&$\lambda_1 g $&$\lambda_3 g$&
$\partial_x,\,\,\dfrac 1g \partial_t,\,\, 3\dfrac Tg\partial_t+\left(x-\lambda_3\varkappa T\right)\partial_x- \left(u+\varkappa\right)\partial_u$\\
\hline
\end{tabular}
\\[2ex]
\parbox{220mm}{Here $g(t)$ is an arbitrary smooth nonvanishing function, $T=\int\! g(t)\, {\rm d}t$; $\lambda_i,$ $i=1,2,3$, $\alpha$, $\beta$ and $\rho$ are arbitrary constants with $\lambda_1\alpha\neq0$, $\rho\neq-1,-\frac43$; $\varkappa=\frac12{\lambda_3}/{\lambda_1}.$ In Case I.1 $(\rho,\lambda_2)\neq(0,0),$ in Case III  $\lambda_3\neq0.$}
\end{center}
\end{sidewaystable}

\begin{theorem}
The kernel of the maximal Lie invariance algebras of equations from class~\eqref{eq_gGardner}
coincides with the one-dimensional algebra $\langle\partial_x\rangle$.
All possible $G^\sim$-inequivalent (resp. $\hat G^\sim$-ine\-qui\-valent) cases of extension of the maximal Lie invariance algebras are exhausted
by Cases I--IV of Table~1 (resp. Table~2).
\end{theorem}

In order to get the most general forms of arbitrary elements of class~\eqref{eq_gGardner} (not simplified by equivalence transformations) we should apply transformation~\eqref{gen_tr} to the equations~\eqref{eq_Gardner} with $k$ and $f$ presented in Table~2 or even simpler transformation~\eqref{gen_tr} with $\delta_3=0$ to the equations~\eqref{eq_Gardner} with $k$ and $f$ presented in Table~1. Then the same transformations should be applied to the corresponding Lie symmetry generators. We present the obtained results in Table~3.

\subsection{Group classification via mapping between classes}

Usually it is easier to solve the group classification problem for a class that is normalized in the usual sense than the group classification problem for a class normalized
in the generalized or generalized extended sense.  The optimal choice for solving
group classification problem for certain classes is the method based on mapping between classes. We would like to show briefly how this method works using the example of variable coefficient Gardner equations.

Consider the family of transformations parameterized by the arbitrary elements $f$, $g$ and $k$ of the class~\eqref{eq_gGardner},
\begin{equation}\label{eq_tr_FL}
\tilde t=\int\!\! g(t)\, {\rm d}t,\qquad \tilde x=x+\int\! \frac{k(t)^2}{4f(t)}\,{\rm d}t,\qquad \tilde u=u+\frac {k(t)}{2f(t)}.
\end{equation}
This family of transformations maps class~\eqref{eq_gGardner} to the class of variable coefficient mKdV equations with forcing term (tildes are skipped),
\begin{equation}\label{fmkdv}
u_t+F(t)u^2u_x+u_{xxx}=L(t),\quad F\neq0,
\end{equation}
where arbitrary elements $F$ and $L$ are expressed via $f$, $g$ and $k$ as
\begin{equation}\label{eq_FL}
F(\tilde t)=\frac {f(t)}{g(t)},\qquad L(\tilde t)=\frac1{2g(t)}\left(\frac {k(t)}{f(t)}\right)_t.
\end{equation}
Similarly to~\eqref{eq_gGardner} the class~\eqref{fmkdv} is also a subclass of the normalized class~\eqref{EqvcmKdV}. So, we can easily deduce its equivalence group from the equivalence group of~\eqref{EqvcmKdV}.  The following assertion is true.
\begin{theorem}
The class~\eqref{fmkdv} is normalized in the usual sense. The usual equivalence group~$G^\sim_2$ of this class is formed by the transformations
\begin{gather*}\tilde t={\delta_1}^3t+\delta_0,\quad \tilde x=\delta_1x+\delta_3,\quad
\tilde u=\delta_2 u, \quad
\tilde F(\tilde t)=\frac{F(t)}{{\delta_1}^2{\delta_2}^2},\quad\tilde L(\tilde t)=\frac{\delta_2}{{\delta_1}^3}L(t),
\end{gather*}
where  $\delta_i$, $i=0,\dots,3,$ are arbitrary constants with
$\delta_1\delta_2\not=0$.
\end{theorem}

Using the classical Lie symmetry method described in previous section we find that the general form of the infinitesimal generator
is $Q=(3c_1t+c_0)\partial_t+(c_1x+c_2)\partial_x+c_3u\partial_u,$
where $c_i,$ $i=0,\dots,3$, are arbitrary constants. The classifying equations have the form
\begin{gather*}
(3c_1 t+c_0) F_t=-2(c_1+c_3)F,\quad (3c_1 t+c_0) L_t=(c_3-3c_1)L.
\end{gather*}
Integrating these equations up to $G^\sim_2$-equivalence we get the complete group classification of the class~\eqref{fmkdv}. The results are summarized in the following statement.
\begin{theorem}
The kernel of the maximal Lie invariance algebras of equations from class~\eqref{fmkdv}
coincides with the one-dimensional algebra $\langle\partial_x\rangle$.
All possible $G^\sim_2$-inequivalent  cases of extension of the maximal Lie invariance algebras are exhausted
by the cases:
\smallskip

{\rm I.} $F=\lambda_1 t^\rho$, $L=\delta t^{-\frac{3\rho+8}{6}}\colon\quad$ $A^{\rm max}=\left\langle\partial_x,\,3t\partial_t+x\partial_x-\frac{3\rho+2}{2}  u\partial_u\right\rangle$;

\smallskip

{\rm II.} $F=\lambda_1 e^{t}$, $L=\delta e^{-\frac 12 t}\colon\quad$
$A^{\rm max}=\langle\partial_x,\,2\partial_t-u\partial_u\rangle$;

\smallskip

{\rm III.}  $F=\varepsilon $, $L=1 \colon\quad$
$A^{\rm max}=\langle\partial_x,\,\partial_t\rangle$;

\smallskip

{\rm IV.}  $F=\varepsilon $, $L=0 \colon\quad$
$A^{\rm max}=\langle\partial_x,\,\partial_t,\, 3t\partial_t+x\partial_x- u\partial_u\rangle$.

\smallskip

\noindent
Here $\lambda_1$ and $\rho$ are arbitrary constants with $\lambda_1\neq0$, $\delta\in\{0,1\}$, $\varepsilon=\pm1$. In Case I $(\rho,\delta)\neq(0,0)$.
\end{theorem}

In Table~4 we present the group classification of class~\eqref{fmkdv}, where the forms of arbitrary elements are not simplified by equivalence transformations.
It is easy to see that though classes~\eqref{eq_gGardner} and~\eqref{fmkdv} are similar with respect to point transformation, the group classification of the latter class is much simpler (cf. Tables~3 and~4).

\begin{table}[t!]\small \renewcommand{\arraystretch}{1.7}
\centering
\setcounter{tbn}{-1}
\refstepcounter{table}\label{TableLieSym4}
\textbf{Table~\thetable.}
The complete list of Lie symmetry extensions for the class~\eqref{fmkdv}.
\\[2ex]
\begin{tabular}{|c|c|c|l|}
\hline
no.&$F(t)$&$L(t)$&\hfil Basis of $A^{\max}$ \\
\hline
0&$\forall$&$\forall$&$\partial_x$\\
\hline
I&
$\lambda_1 (\alpha t+\beta)^\rho$&$\lambda_2 (\alpha t+\beta)^{-\frac{3\rho+8}{6}}$&$\partial_x,\,3(\alpha t+\beta)\partial_t+\alpha x\partial_x-\frac{3\rho+2}{2} \alpha u\partial_u$\\
\hline
II&$\lambda_1 e^{\alpha t}$&$\lambda_2 e^{-\frac 12\alpha t}$&
$\partial_x,\,2\partial_t-\alpha u\partial_u$\\
\hline
III &$\lambda_1 $&$\lambda_2$&
$\partial_x,\,\partial_t$\\
\hline
IV&$\lambda_1 $&$0$&
$\partial_x,\,\partial_t,\, 3t\partial_t+x\partial_x- u\partial_u$\\
\hline
\end{tabular}
\\[2ex]
\parbox{130mm}{Here $\lambda_1$, $\lambda_2$, $\alpha$, $\beta$ and $\rho$ are arbitrary constants with $\lambda_1\alpha\neq0$. In Case~I $(\rho,\lambda_2)\neq(0,0)$, in Case~III $\lambda_2\neq0.$}
\end{table}

Equations from class~\eqref{eq_gGardner} with the coefficients presented in the cases of Tables~1--3 are linked with the equations~\eqref{fmkdv} having coefficients presented in the case denoted by the same Roman numeral in Table~4 via point transformations.
We could obtain Table~3  directly from Table~4 in the way shown in the following examples.
\begin{example}
Consider Case~IV of Table~4. We substitute the corresponding values of the functions $F$ and $L$ to~\eqref{eq_FL}, this results in the equations
$f/g=\lambda_1$, $\left(k/f\right)_t=0$. The general solution is given by $f=\lambda_1g,$ $k=\lambda_3g$ (we use notation $\lambda_3$ for the integration constant to be able to compare easily the results obtained by the method based on mapping between classes with those obtained by equivalence based approach). Then the transformation~\eqref{eq_tr_FL} takes the form
\[\tilde t=\int\!\! g(t)\, {\rm d}t,\quad \tilde x=x+\frac{\lambda_3^2}{4\lambda_1}\int\!\! g(t)\, {\rm d}t,\quad \tilde u=u+\frac {\lambda_3}{2\lambda_1}.\]
Under the action of this transformation differential operators are transformed as follows $\partial_{\tilde t}=\frac1g\partial_t-\frac{\lambda_3^2}{4\lambda_1}\partial_x,$ $\partial_{\tilde x}=\partial_x$ and $\partial_{\tilde u}=\partial_u.$ Therefore, the operators presented in Case~IV of Table 4 take the form $\partial_x$, $\frac1g\partial_t-\frac{\lambda_3^2}{4\lambda_1}\partial_x$ and $3\frac {\int\!\! g(t)\, {\rm d}t}g\partial_t+\left(x-\frac{\lambda_3^2}{2\lambda_1}\int\!\! g(t)\, {\rm d}t\right)\partial_x- \left(u+\frac{\lambda_3}{2\lambda_1}\right)\partial_u$, respectively. It is convenient to choose a basis of Lie symmetry algebra spanned by these operators as presented in Case~IV of Table~3.
\end{example}
\begin{example}
Consider Case~I of Table~4 extended by the equivalence transformations from~$G^\sim_2$, i.e., the case $F=\lambda_1 (\alpha \tilde t+\beta)^\rho$ and  $L=\lambda_2 (\alpha \tilde t+\beta)^{-\frac{3\rho+8}{6}}.$ As $\tilde t=\int\! g(t)\, {\rm d}t$, equations~\eqref{eq_FL} imply $f(t)=\lambda_1 g(t)(\alpha \int\! g(t)\, {\rm d}t+\beta)^\rho$ and $(k/f)_t=2\lambda_2 g(t)(\alpha \int\! g(t)\, {\rm d}t+\beta)^{-\frac{3\rho+8}{6}}$. If we denote $\int\! g(t)\, {\rm d}t$ by $T$, then $k(t)=2\lambda_2 f(t)\int\,(\alpha T+\beta)^{-\frac{3\rho+8}{6}} {\rm d}T$. Finally, \[k(t)=\begin{cases}2\lambda_2\lambda_1 g(t)(\alpha T+\beta)^\rho\left(-\frac{6}{\alpha(3\rho+2)}(\alpha T+\beta)^{-\frac{3\rho+2}{6}}+\lambda_3\right),&\mbox{if}\quad \rho\neq-2/3, \\ 2\lambda_2\lambda_1 g(t)(\alpha T+\beta)^\rho\left(\frac{1}{\alpha}\ln|\alpha T+\beta|+\lambda_3\right),&\mbox{if}\quad \rho=-2/3.\end{cases}\]
After redenoting the constants $\lambda_i$, $i=1,2,3,$ it is easy to see that we get Cases I.1 and I.2 of Table~3, respectively. To obtain the corresponding symmetry operators one  should make the change of variables~\eqref{eq_tr_FL} in the operators $X_1=\partial_{\tilde x}$ and $X_2=3(\alpha \tilde t+\beta)\partial_{\tilde t}+\alpha \tilde x\partial_{\tilde x}-\frac{3\rho+2}{2} \alpha {\tilde u}\partial_{\tilde u}$. As transformation~\eqref{eq_tr_FL} contains integral of $k^2/(4f)$ the special cases of integration arise for $\rho=-1$ and $\rho=-4/3$. These cases correspond to Cases I.3 and I.4 in Table~3, that do not differ by forms of $f$ and $k$ from the general Case I.1 but differ from this case essentially by forms of the second symmetry generator.
\end{example}

In a similar way the other cases of Table~3 can be derived.

\noprint{
It is interesting to note that though transformation~\eqref{eq_tr_FL} looks quite complicated it preserves the form of  differential operators presented in Cases~I.1, II and IV of Tables~2 and~4.
For example, the equation $u_t+\lambda_2t^{\frac{\rho}2-\frac13}uu_x+\lambda_1t^\rho u^2u_x+u_{xxx}=0$ (Case~I.1 of Table~2) is similar to the equation $u_t+\lambda_1t^\rho u^2u_x+u_{xxx}=-\frac{\lambda_2}{12\lambda_1}(3\rho+2)t^{\frac{3\rho+8}{6}}$ (Case~I of Table~4) with respect to the point transformation
\[\tilde t=t,\quad\tilde x=x+\frac{3\lambda_2^2}{4\lambda_1}t^{\frac13},\quad\tilde u=u+\frac{\lambda_2}{2\lambda_1}t^{-\frac{3\rho+2}6}.\]
It can be verified by direct calculations that this transformation leaves the Lie symmetry operators $\partial_x$ and $3t\partial_t+x\partial_x-\frac{3\rho+2}{2} u\partial_u$ invariant.}

\section*{Conclusion}
We have presented two alternative ways to completely solve the group classification problem for class~\eqref{eq_gGardner}.
These are the gauging of arbitrary elements by equivalence transformations and the mapping of the initial class~\eqref{eq_gGardner} to the similar class~\eqref{fmkdv} of simpler structure. The advantage of the first approach is that it is fully algorithmic, in contrast to the second one which requires some guessing.
Nevertheless, for some classes the method of mapping between classes seems to be the unique opportunity to get the exhaustive group classification~\cite{VPS2009}.

We have found that, besides the usual equivalence group~$G^\sim$, class~\eqref{eq_gGardner} admits the wider generalized extended equivalence group~$\hat G^\sim$.
Although the exhaustive group classification of class~\eqref{eq_gGardner} can be achieved even using the usual equivalence group, we have shown that the generalized extended equivalence group provides more simplification and allows one to write down the classification list in a simple and concise form (compare Table~1 with Table~2). The most general forms of arbitrary elements that provide Lie symmetry extensions for the corresponding equations from class~\eqref{eq_gGardner} can be derived then applying equivalence transformations (Table~3).  Obviously, the use of the widest possible equivalence group is preferable for solving  group classification problems, but the worst choice is to neglect opportunity of utilizing equivalence transformations at all.

If we compare now the results
derived in~\cite{Molati&Ramollo2012} with those adduced in Table~3 it is easy to see that the cases of Lie symmetry extension presented in~\cite{Molati&Ramollo2012} are several particular specifications of Cases I--III from Table~3 for certain fixed values of the function~$g$. Furthermore the case of the three-dimensional maximal Lie symmetry algebra was not indicated in~\cite{Molati&Ramollo2012}.

\bigskip\par\noindent{\bf Acknowledgements.}
This work was partially supported by a grant from the Niels Henrik Abel Board.
OV would like to thank the University of Cyprus for hosting her during this project.
The authors express their gratitude to Roman Popovych for useful discussions.

\end{document}